# Highly in-plane anisotropic optical properties of fullerene monolayers

Danwen Yuan[1,2#], Hanqi Pi[3,4#], Yi Jiang[3,4#], Yuefang Hu[1,2], Liqin Zhou[3,4], Yujin Jia[3,4], Gang Su[5], Zhong Fang[3,4], Hongming Weng[3,4,6*], Xinguo Ren[3,4,6*], Wei Zhang[1,2*]

[1] *Fujian Provincial Key Laboratory of Quantum Manipulation and New Energy Materials, College of Physics and Energy, Fujian Normal University, Fuzhou 350117, China*

[2] *Fujian Provincial Collaborative Innovation Center for Advanced High-Field Superconducting Materials and Engineering, Fuzhou 350117, China*

[3] *Beijing National Laboratory for Condensed Matter Physics, Institute of Physics, Chinese Academy of Sciences, Beijing 100190, China*

[4] *University of Chinese Academy of Sciences, Beijing 100049, China*

[5] *Kavli Institute for Theoretical Sciences and CAS Center for Excellence in Topological Quantum Computation, University of Chinese Academy of Sciences, Beijing 100190, China*

[6] *Songshan Lake Materials Laboratory, Dongguan, Guangdong 523808, China*

[#]Danwen Yuan, Hanqi Pi, and Yi Jiang contributed equally.
[*]Corresponding Author:
Hongming Weng email: hmweng@iphy.ac.cn;
Xinguo Ren email: renxg@iphy.ac.cn;
Wei Zhang email: zhangw721@163.com.

**ABSTRACT:** Both the intrinsic anisotropic optical materials and fullerene-assembled 2D materials have attracted a lot of interests in fundamental science and potential applications. The synthesis of a monolayer (ML) fullerene makes the combination of these two features plausible. In this work, using first-principles calculations, we systematically study the electronic structure, optical properties of quasi-hexagonal phase (qHP) ML and quasi-tetragonal phase (qTP) ML fullerenes. The calculations of qHP ML show that it is a semiconductor with small anisotropic optical absorption, which agrees with the recent experimental measurements. However, the results for qTP ML reveal that it is a semimetal with highly in-plane anisotropic absorption. The dichroic ratio, namely the absorption ratio of *x*- and *y*-polarized light $\alpha_{xx}/\alpha_{yy}$, is around 12 at photon energy of 0.29 eV. This anisotropy is much more pronounced when the photon energy is between 0.7 and 1.4 eV, where $\alpha_{xx}$ becomes nearly zero while $\alpha_{yy}$ is more than two orders of magnitude larger. This indicates qTP ML as a candidate for long-pursuit lossless metal and a potential material for atomically thin polarizer. We hope this will stimulate further experimental efforts in the study of qTP ML and other fullerene-assembled 2D materials.

## INTRODUCTION

Stimulated by the successful synthesis of a monolayer (ML) fullerene network very recently,[1] there have been renewed and instant interests in the study of $C_{60}$-based two-dimensional (2D) materials.[2,3] Carbon allotropes constructed of *sp*, *sp*$^2$, and *sp*$^3$ hybrid bonds have remarkable structures of different dimension and rich properties in both physics and chemistry.[4-10] Most famous examples include zero-dimensional (0D) carbon cluster fullerene,[11,12] 1D carbon nanotube,[13,14] 2D atom-thick graphene[4,7] and 3D diamond.[15-18] In addition to carbon atoms, clusters like $C_{60}$ fullerene have also acted as a very promising building block, which can form novel nanocluster-assembled materials with controllable synthesis and tailorable characteristics.[19-24] It is of significance to find more $C_{60}$ assembled stable structures and investigate in detail their electronic structure and physical properties. This has been especially driven by the $C_{60}$-based MLs discovered recently.[1,25-27] These MLs are helpful to illustrate the mechanism of clusters evolving into crystals, and shed light on designing another new class of nanocluster-inspired materials with unusual properties.[28-30] For example, the bands composed by molecular orbitals are remarkably narrow and the correlation effect in them is enhanced, which may lead to Mott insulator, superconductivity and nontrivial topological states.[31-33]

In this paper, we have calculated the electronic structure, optical properties and scanning tunneling microscopy (STM) image for both quasi-hexagonal phase (qHP) and quasi-tetragonal phase (qTP) MLs reported in recent experimental work.[1] For qHP ML, the theoretical calculations are consistent with the experimental reports. For the qTP ML, although it has not been finally identified in experiments, we predict that it might be a semimetal with significant in-plane anisotropic optical properties, being quite different from qHP ML. The real-space and momentum-space pictures for understanding these polarization-dependent optical properties have been established based on the elementary band representation (EBR) analysis[34] and selection rules for electric dipole transition. These results may inspire the potential experimental realization of qTP ML, which is very promising for applications in 2D electronic and optical devices, such as the lossless metal[35] and sub-nanometer coating material for polarizer.

**METHODS**

Using Vienna *ab initio* simulation package (VASP),[36] we perform the first-principles calculations within density functional theory (DFT). We use the generalized gradient approximation (GGA) in the form of the Perdew, Burke, and Ernzerhof to deal with the exchange-correlation functional.[37] The kinetic energy cutoff for the plane wave expansion is set to be 500 eV. We use $7 \times 13 \times 7$, $15 \times 9 \times 1$, $9 \times 9 \times 9$ and $9 \times 9 \times 1$ k-point meshes in self-consistent electronic structure calculations for the 3D bulk of qHP, 2D ML of qHP, 3D qTP bulk, and qTP ML, respectively. The vacuum region is set to about 20 Å in the supercell of qHP ML and qTP ML to avoid the interaction among neighboring MLs. The crystal structures of the bulk qHP and qTP are taken from experimental reports. The MLs are directly exfoliated from their bulk phases thanks to the layered structure. To remedy the problem that the PBE functional generally underestimates the band gap, we also perform the HSE06 hybrid functional calculations[38,39] using all-electron electronic-structure code FHI-aims.[40-42] The FHI-aims-2020 numeric atom-centered orbitals with "tight" setting are used in all calculations. Moreover, the optical properties calculations are also executed using PBE and HSE06 methods, respectively. Finally, the STM images are simulated in the constant height mode, according to the Tersoff-Hamann approximation.[43]

**RESULTS AND DISCUSSION**

**Crystal structure.** The qHP and qTP bulk single crystals of polymeric $C_{60}$ have been synthesized in experiment.[1] Figure 1 shows the crystal structures of these two materials. The Mg ions in bulk phases are not shown and have not been included in the calculations in this work. The ML phases have no Mg ions and can be neutralized according to the experimental description. The qHP bulk single crystal is in space group *Pc* (No. 7) and the lattice parameters from single-crystal x-ray diffraction (XRD) measurement are a = 17.570 Å, b = 9.169 Å, c = 16.001 Å. The $C_{60}$ cages form a 2D periodic network stacking along the a-axis direction in ABAB... order, as shown in Figure 1(a). The corresponding ML has been exfoliated from the bulk single crystal. The crystal structure of qHP ML is displayed in Figure 1(c), and it maintains the same space group as the bulk one. As illustrated in Figure 1(c), each $C_{60}$ is connected with six surrounding $C_{60}$ fullerenes, which are approximately located at the hexagonal positions. Four adjacent $C_{60}$ clusters are connected with the central $C_{60}$ by four $sp^3$-like C-C bonds along ±30° off the a-axis (*x* direction). The other two are connected by two neighboring $sp^3$-like C-C bonds, which forms a square ring along the b-axis (*y* direction). Figure 1(b) depicts the crystal structure of the qTP bulk single crystal, whose space group is *C2/m* (No. 12) and the lattice constants observed in the experiment are a = 9.308 Å, b = 9.032 Å, c = 14.780 Å. Like the qHP bulk structure, the qTP bulk single crystal also has the layered structure, in which the polymeric $C_{60}$ fullerene layers are stacking along the c-axis in ABAB...order. The possible exfoliated qTP ML structure is shown in Figure 1(d) and the space group becomes *P2/m* (No. 10). As exhibited in Figure 1(d), the adjacent $C_{60}$ clusters forming a square lattice are connected by two C-C single bonds along the a-axis and two square rings along the b-axis. This asymmetric bond arrangement in qHP ML and qTP ML makes them inequivalent along a and b axes, resulting in the anisotropy of these two MLs.

**Charge difference density.** To analyze the bonding properties, we perform the charge difference density (CDD) around the C-C bridge bonds linking the $C_{60}$ clusters in qHP ML and qTP ML. The calculated charge difference density is defined as: $\Delta\rho = \rho(total) - \rho(C)$, where the $\rho(total)$ is the charge density of the whole system, and the $\rho(C)$ represents the summation of charge density of individual C atoms. CDD distributions around the C-C bridge bonds in qHP ML are displayed in Figures 2(a) and (b). From the pictures, we can see that the central C atoms in the red circle form 4 bonds with adjacent C atoms, and yellow charge density means the electrons accumulate around C-C bond center. Thus, these 5 atoms construct a tetrahedron as in the case of diamond. This means that the C-C bridging bonds form very strong $sp^3$-like covalent bonds, which may lead to the stable structure of qHP ML. However, bond angles related to the central fourfold-coordinated atoms are different from the perfect $sp^3$ hybridization angle of 109.47°.[25] The bond angles of $sp^3$-like atoms here are 93.57°, 105.61°, 106.28°, 113.79°, 118.18°, and 118.39°. In addition, the length of the inter-cluster bond is 1.449 Å for bonds in square ring and 1.589 Å for the single bond, which are comparable to the bond length of ideal $sp^2$ bond in graphene ~1.42 Å and $sp^3$ bond in diamond ~1.548 Å.[12] This also indicates the strength of C-C bridge bonds is very strong, which may help understand the stability of qHP ML synthesized in experiment. Such $sp^3$-like fourfold-coordinated C atoms also appear in the bridging bonds in qTP ML, as shown in Figures 2(c) and (d), respectively. This may indicate the crystal structure of qTP ML is also very stable, and very promising for experimental synthesis and observation.

**Band structure.** The band structures of qHP bulk, qTP bulk and related MLs using PBE exchange-correlation functional are calculated and presented in Figure 3. As shown in Figure 3(a), the valence band maximum (VBM) of the qHP bulk single crystal is at the $Y_2$ point, and the conduction band minimum (CBM) is at the Γ point, forming an indirect energy gap of 0.20 eV. The direct band gap at Γ point is 0.20 eV, showing the semiconducting features. Accordingly, Figure 3(b) shows the band structure of the qHP ML. The VBM and CBM are at the Γ and Y points respectively, resulting in an indirect gap of 0.56 eV. The direct band gap at Γ point is 0.59 eV, which is larger than that of the qHP bulk single crystal due to the missing of interlayer interactions. Like the qHP bulk single crystal, the qHP ML also shows semiconducting characteristic. The PBE band structures of the bulk and ML of qTP are also studied. It can be seen from Figure3(c) that there are energy bands crossing the Fermi level along the Γ-C direction, suggesting that the qTP bulk exhibits metallicity. For the ML qTP, Figure 3(d) shows the typical semimetal feature with both electron and hole pockets.

Since the PBE functional generally underestimates the band gap of semiconductor and to obtain the band gaps of these two MLs accurately, the hybrid functional HSE06 is employed. The results are shown in Figure 4. The indirect gap of qHP ML increases to 0.99 eV, and the direct gap at Γ point increases to 1.07 eV, which is smaller than that of experiment value ~1.6 eV estimated from optical measurements. As for the qTP ML, the band structure is altered by HSE06 and it is more like a narrow gap semiconductor. Although both qHP ML and qTP ML are $C_{60}$-assembled ML and the adjacent $C_{60}$ clusters are connected by similar $sp^3$-like C-C bonds, they show quite different electronic band structure and optical properties. This indicates the design of various $C_{60}$-assembled 2D materials could have plenty of properties and wide applications.

**Anisotropic optical property.** To investigate the optical properties, we calculate the optical absorption spectra of qHP ML and qTP ML within both PBE and HSE06 functionals. As displayed in Figure 5(a), the absorption $\alpha_{xx}$ and $\alpha_{yy}$ curves of qHP ML display the transition threshold at about 0.4 eV. The $\alpha_{xx}$ absorption spectrum has first two peaks at 0.81 eV and 1.17 eV for the incident light polarized along *x* direction. The $\alpha_{yy}$ absorption has one main peak around 1.0 eV. The dichroic ratio, the ratio of the absorptions of lights with in-plane polarization at perpendicular directions, of qHP ML around this peak is estimated to be $\alpha_{xx}/\alpha_{yy} = 0.43$ or $\alpha_{yy}/\alpha_{xx} = 2.35$. This ratio is slightly larger than that estimated from conductivity measurement.[1] The in-plane anisotropic optical absorption is consistent with the asymmetric lattice structure, in which the connectivity of $C_{60}$ clusters along *x* and *y* is different. In contrast, the anisotropy in the absorption spectra of qTP ML is much larger than qHP ML, e. g., the dichroic ratio at the first absorption peak of 0.29 eV is $\alpha_{xx}/\alpha_{yy} = 12$. This is unexpected since the difference in a- and b-lattice constant of qTP ML is less than that of qHP ML. Due to the semimetal feature and quite small electron and hole pockets, we ignore the Drude type intraband contribution to the absorption in photon energy higher than 0.2 eV. We mainly discuss two distinct features of the anisotropic absorption spectra in Figure 5(b). One is those in the photon energy ranging from 0.2 eV to 0.6 eV, the other is in the region from 0.7 eV to 1.4 eV. $α_{xx}$ is much larger than $α_{yy}$ in the former region and nearly zero in the latter one. This nearly zero absorption of *x*-polarized light makes qTP ML a potential candidate for long pursuit lossless metal.[35] Another possible application is to be used as a sub-nanometer coating material for polarizer of light in this energy region. We further compute the optical absorption spectra of qHP ML and qTP ML using the HSE06 functional to overcome the shortcomings of PBE functional. As shown in Figures 5 (c) and (d), the major correction of HSE06 to the absorption curves is to shift them to higher energy and this might be closer to the experimental measurements. For qTP ML, the nearly zero absorption region of $α_{xx}$ is shifted to 1.0 and 2.0 eV, which covers part of infrared and visible lights. To understand the mechanism underlying these intriguing optical properties in qTP ML, we have proposed the real space and momentum space pictures in the following.

**IRREP analysis of qTP ML.** We analyze the irreducible representations (IRREPs) for the bands of qTP ML. The IRREPs[46] at high-symmetry points (HSPs) and corresponding elementary band representations (EBRs)[34,47] of the bands near the Fermi level in qTP ML are summarized in Table 2. Note that the (N − 4)-th to (N − 1)-th bands are entangled together and do not have a unique EBR decomposition. It can be seen that Wyckoff positions (0, 0.5) and (0.5, 0.5) must be occupied by electronic Wannier centers and there are no ionic occupations, which means qTP ML is an obstructed atomic insulator (OAI) with separated ion and electron charge center.[44,45] These two Wyckoff positions are the center of $C_{60}$ clusters and the center of the single bonds in qTP ML, respectively. The mismatch of Wannier centers and ionic positions at (0.5, 0.5) can be directly seen from the CDD in Figure 2(d). However, the Wannier center at the center of $C_{60}$, as a result of charge averaging, cannot be seen directly from the charge distribution.

We use the IRREPs to analyse the absorption spectra of qTP ML. As illustrated in Figure 5(b), the absorption spectra of $\alpha_{xx}$ and $\alpha_{yy}$ both peak at 0.29 eV, but have strong anisotropy, i.e., the peak value of $\alpha_{xx}$ is nearly twelve times of $\alpha_{yy}$. $\alpha_{xx}$ is nearly zero in the range 0.7~1.4 eV, while $\alpha_{yy}$ has finite value and several peaks in this energy range. To figure out the reason for the anisotropy, we first give a real-space picture based on EBRs, and then perform a detailed analysis of the electron transitions in the BZ, i.e., the momentum space.

The anisotropic peaks at 0.29 eV can be understood from a simple real-space picture using EBRs. The N-th and the (N + 1)-th bands are the highest occupied molecular orbital (HOMO) and the lowest unoccupied molecular orbital (LUMO). From the EBR decomposition, they can be seen as the $B_u$ and $A_g$ orbitals at the same Wyckoff position (0.5, 0.5), respectively. From Table 1, the IRREPs $B_u$ and $A_g$ can be generated by $p_x$ and $s$ orbital, respectively. As a result, the HOMO and LUMO form isolated flat valence and conduction bands, respectively. The energy difference between them is about 0.29 eV and can be approximated as isolated $p_x$ and $s$ orbitals placed at the same Wyckoff position (0.5, 0.5), respectively. This leads to strong $x$-polarized optical absorption and much weak absorption for $y$-polarized light.

We further analyze the electron transitions of qTP ML allowed by dipole radiation selection rules[48] in the BZ. In Figure 3(d), we mark the energy bands at the HSPs with their IRREPs in $C_{2h}$ group. As the high-symmetry lines Γ-Y and X-M have the little group $C_2$, the plus symbol (+) and the minus symbol (−) denote the energy bands belonging to the IRREPs $A$ and $B$, respectively. Besides, a polar vector such as the momentum operator $\hat{p}$ in $C_2$ group transforms as $A$ for component $\hat{p}_y$ and as $B$ for components $\hat{p}_x$ and $\hat{p}_z$, as shown in Table 1. It implies that the selection rules are different for light polarized along and transverse to the two-fold rotation axis. The transitions between states $|i\rangle$ and $|j\rangle$ excited by light polarized in $k$ direction are allowed if the representation of dipole transition matrix element $\langle i|\hat{p}_k|j\rangle$ ($k$ = $x$, $y$, $z$) contains the identity representation. On high-symmetry lines, symmetries cast selection rules which may force transition momentum to be zero. However, for generic momentum, there is no symmetry constraint and the transition momentum is generally nonzero.

Therefore, along Γ-Y and X-M paths, electron dipole transitions are allowed between bands with different representations for radiation polarized in *x* direction and bands with the same representations for polarization in *y* direction. Similarly, along Γ-X and Y-M paths belonging to the little group $C_s$, radiation in *x* direction connects bands with the same representation while radiation in *y* direction has the opposite selection rules. The red arrows in Figure 3(d) indicate the electron transitions excited by photon at 0.29 eV. According to the selection rules, all of them could be induced by radiation in *x* direction but only transitions in M-Γ path could be excited by radiation in *y* direction. This explains the strong anisotropy of the absorption spectrum and the higher intensity of $\alpha_{xx}$ at 0.29 eV

To understand the nearly zero $\alpha_{xx}$ at the photon energy around 1.0 eV, we illustrate the distribution of dipole transition momentum strength $D_k \equiv |\langle i|\hat{p}_k|j\rangle|^2$ ($k = x, y$) in the BZ in Figure 6, in which the transition energy is sampled from 0.95 to 1.05 eV. It is obvious that $D_y$ has a larger nonzero area and is averagely two orders of magnitude greater than $D_x$. The corresponding electron transitions on the high-symmetry lines are denoted in Figure 3(d) with blue ($D_y$) and green ($D_x$) arrows. Therefore, after integration of the dipole transition momentum strength over the whole BZ weighted by the joint density of states, the absorption of $\alpha_{xx}$ is much smaller than $\alpha_{yy}$ in this region.

**STM image.** We perform the first-principles calculations to simulate the STM images of qHP ML and qTP ML to investigate their surface topography. The STM image of qHP ML is shown in Figure 7(a). The bright C atoms are arranged in an ordered structure, which looks like an ant. However, the upper side and lower side of the "ant" are not symmetric. Comparing the STM image with the crystal structure of qHP ML shown in Figures 7(b) and (c), it's clear that the STM image highlights the [6-5] bond between the five-member ring (5-MR) and 6-MR on the surface of $C_{60}$ nanocluster. The STM image of qTP ML displayed in Figure 7(d) shows a relatively bright pattern according to its semi-metallic nature. The two brightest spots in the STM image correspond to the [6-5] bond between 5-MR and 6-MR on the surface of $C_{60}$ cluster, while two second-brightest spots relate to the [6-6] bond between two 6-MRs, as shown in Figures 7(e) and (f). These theoretical STM images can provide fingerprints of two MLs for future experimental investigations.

**CONCLUSION**

In this paper, the electronic and optical properties of qHP ML and qTP ML are obtained with PBE exchange-correlation functional and HSE06 hybrid functional calculations. The calculated results show that qHP ML is a semiconductor and has marginal optical anisotropy with dichroic ratio around 2.34, being consistent with the experimental measurements. On the contrary, qTP ML is predicted to be a semimetal with strong anisotropic optical absorption. In the infrared and visible light region, the *x*-polarized light is nearly absorptionless while the absorption of *y*-polarized light is more than two orders of magnitude larger. The underlying physics for this anisotropy has been revealed in both real-space and momentum space pictures based on the elementary band representation analysis and selection rules constrained by symmetry in electronic dipole transition. Finally, the STM images are simulated to guide the experimental identification of both qHP and qTP MLs. We think the intrinsic high in-plane anisotropy makes the qTP ML very promising for optoelectronic devices, such as the candidate for lossless metal and sub-nanometer polarizer.

**Notes**

The authors declare no competing financial interest.

**ACKNOWLEDGMENT**

The authors thank the valuable discussion with Daoben Zhu, Ling Lu and Jian-Tao Wang. This work was supported by the National Natural Science Foundation of China

**REFERENCES**

(1) Hou, L.; Cui, X.; Guan, B.; Wang, S.; Li, R.; Liu, Y.; Zhu, D.; Zheng, J. Synthesis of a monolayer fullerene network. *Nature* **2022**, 606, 507-510.

(2) Yu, L.; Xu, J.; Peng, B.; Qin, G.; Su, G.; Flat electronic band structure and anisotropic properties of two-dimensional fullerene networks. 6 Jul 2022. arXiv. https://arxiv.org/abs/2207.02781.

(3) Tromer, R. M.; Junior, L. A.; Galvão, D. S.; A DFT Study of the Electronic, Optical, and Mechanical Properties of a Recently Synthesized Monolayer Fullerene Network. 4 Jul 2022. arXiv. https://arxiv.org/abs/2207.01663.

(4) Geim, A. K.; Novoselov, K. S. The rise of graphene. *Nat. Mater.* **2007**, 6, 183-191.

(5) Novoselov, K. S.; Geim, A. K.; Morozov, S. V.; Jiang, D.; Zhang, Y.; Dubonos, S. V.; Grigorieva, I. V.; Firsov, A. A. Electric field effect in atomically thin carbon films. *Science* **2004**, 306, 666-669.

(6) Fan, Q.; Yan, L.; Tripp, M. W.; Krejˇcí, O.; Dimosthenous, S.; Kachel, S. R.; Chen, M.; Foster, A. S.; Koert, U.; Liljeroth, P.; Gottfried, J. M. Biphenylene network: A nonbenzenoid carbon allotrope. *Science* **2021**, 372, 852-856.

(7) Hirsch, A. The era of carbon allotropes. *Nat. Mater.* **2010**, 9, 868-871.

(8) Kolmer, M.; Steiner, A. K; Izydorczyk, I.; Ko, W.; Engelund, M.; Szymonski, M.; Li, A. P; Amsharov, K. Rational synthesis of atomically precise graphene nanoribbons directly on metal oxide surfaces. *Science* **2020**, 369, 571-575.

(9) Qian, L.; Xie, Y.; Zou, M.; Zhang, J. Building a Bridge for Carbon Nanotubes from Nanoscale Structure to Macroscopic Application. *J. Am. Chem. Soc.* **2021**, 143, 18805−18819.

(10) Dunk, P. W.; Kaiser, N. K.; Mulet-Gas, M.; Rodríguez-Fortea, A.; Poblet, J. M.; Shinohara, H.; Hendrickson, C. L.; Marshall, A. G.; Kroto, H. W. The Smallest Stable Fullerene, M@C28 (M = Ti, Zr, U): Stabilization and Growth from Carbon Vapor. *J. Am. Chem. Soc.* **2012**, 134, 9380−9389.

(11) Kroto, H. W.; Heath, J. R.; O'Brien, S. C.; Curl, R. F.; Smalley, R. E. $C_{60}$: Buckminsterfullerene. *Nature* **1985**, 318, 162-163.

(12) Blank, V. D.; Buga, S. G.; Dubitsky, G. A.; Serebryanaya, N. R.; Popov, M. Y.; Sundqvist, B. High-pressure polymerized phases of $C_{60}$. *Carbon* **1998**, 36, 319-343.

(13) Iijima, S.; Ichihashi, T.; Single-shell carbon nanotubes of 1-nm diameter. *Nature* **1993**, 363, 603-605.

(14) Iijima, S. Helical microtubules of graphitic carbon. *Nature* **1991**, 354, 56-58.

(15) Yang, H. X.; Xu, L. F.; Fang, Z.; Gu, C. Z.; Zhang, S. B. Bond-counting rule for carbon and its application to the roughness of diamond (001). *Phys. Rev. Lett.* **2008**, 100, 26101.

(16) Prelas, M. A.; Popovici, G.; Bigelow, L. K.; eds. Handbook of industrial diamonds and diamond films. Marcel Dekker, New York, **1998**.

(17) Angus, J. C.; Hayman, C. C. Low-pressure, metastable growth of diamond and "diamondlike" phases. *Science* **1988**, 241, 913-921.

(18) Xu, J.; Yokota, Y.; Wong, R. A.; Kim, Y.; Einaga, Y. Unusual Electrochemical Properties of Low-Doped Boron-Doped Diamond Electrodes Containing $sp^2$ Carbon. *J. Am. Chem. Soc.* **2020**, 142, 2310−2316.

(19) Yu, H.; Xue, Y.; Li, Y. Graphdiyne and its Assembly Architectures: Synthesis, Functionalization, and Applications. *Adv. Mater.* **2019**, 31, 1803101.

(20) Jena, P.; Sun, Q. Super Atomic Clusters: Design Rules and Potential for Building Blocks of Materials. *Chem. Rev.* **2018**, 118, 5755-5870.

(21) Pekker, S.; Jánossy, A.; Mihaly, L.; Chauvet, O.; Carrard, M.; Forró, L. Single-crystalline $(KC_{60})_n$: a conducting linear alkali fulleride polymer. *Science* **1994**, 265, 1077-1078.

(22) Porezag, D.; Pederson, M. R.; Frauenheim, T.; Köhler, T. Structure, stability, and

vibrational properties of polymerized $C_{60}$. *Phys. Rev. B* **1995**, 52, 14963-14970.

(23) Haddon, R. C.; Hebard, A. F.; Rosseinsky, M. J.; Murphy, D. W.; Duclos, S. J.; Lyons, K. B.; Miller, B.; Rosamilia, J. M.; Fleming, R. M.; Kortan, A. R.; Glarum, S. H.; Makhija, A. V.; Muller, A. J.; Eick, R. H.; Zahurak, S. M.; Tycko, R.; Dabbagh, G.; Thiel, F. A. Conducting films of $C_{60}$ and $C_{70}$ by alkali-metal doping. *Nature* **1991**, 350, 320-322.

(24) Long, V. C.; Musfeldt, J. L.; Kamarás, K.; Adams, G. B.; Page, J. B.; Iwasa, Y.; Mayo, W. E. Far-infrared vibrational properties of high-pressure high-temperature $C_{60}$ polymers and the $C_{60}$ dimer. *Phys. Rev. B* **2000**, 61, 13191-13201.

(25) Okada, S.; Saito, S. Electronic structure and energetics of pressure-induced two-dimensional $C_{60}$ polymers. *Phys. Rev. B* **1999**, 59, 1930-1936.

(26) Xu, C.; Scuseria, G. E. Theoretical predictions for a two-dimensional rhombohedral phase of solid $C_{60}$. *Phys. Rev. Lett.* **1995**, 74, 274-277.

(27) Tanaka, M.; Yamanaka, S. Vapor-Phase Growth and Structural Characterization of Single Crystals of Magnesium Doped Two-Dimensional Fullerene Polymer $Mg_2C_{60}$. *Cryst. Growth Des.* **2018**, 18, 3877-3882.

(28) Makarova, T. L.; Sundqvist, B.; Höhne, R.; Esquinazi, P.; Kopelevich, Y.; Scharff, P.; Davydov, V. A.; Kashevarova, L. S.; Rakhmanina, A. V. Magnetic carbon. *Nature* **2001**, 413, 716-718.

(29) Simon, P.; Gogotsi, Y. Materials for electrochemical capacitors. *Nat. Mater.* **2010,** 7, 320-329.

(30) Cao, Y.; Fatemi, V.; Fang, S.; Watanabe, K.; Taniguchi, T.; Kaxiras, E.; Jarillo-Herrero, P. Unconventional superconductivity in magic-angle graphene superlattices. *Nature* **2018**, 556, 43-50.

(31) Gao, S.; Zhang, S.; Wang, C.; Tao, W.; Liu, J.; Wang, T.; Yuan, S.; Qu. G.; Pan, M.; Peng, S.; Hu. Y.; Li, H.; Huang, Y.; Zhou, H.; Meng, S.; Yang, L.; Wang, Z.; Yao, Y.; Chen, Z.; Shi, M.; Ding, H.; Jiang, K.; Li, Y.; Shi, Y.; Weng. H.; Qian, T. Mott insulator state in a van der Waals flat-band compound. 23 May 2022. arXiv. https://arxiv.org/abs/2205.11462.

(32) Zhang, S; Peng, S.; Dai, X.; Weng, H. Topological States in Chevrel Phase Materials from First-principle Calculations. 15 May 2022. arXiv. https://arxiv.org/abs/2205.07281.

(33) Zhang, Y.; Gu, Y.; Weng. H.; Jiang, K.; Hu, J. Mottness in 2-dimensional van der Waals $Nb_3X_8$. 4 Jul 2022. arXiv. https://arxiv.org/abs/2207.01471.

(34) Cano, J.; Bradlyn, B.; Wang, Z.; Elcoro, L.; Vergniory, M. G.; Felser, C.;Aroyo, M. I.; Bernevig, B. A. Building blocks of topological quantum chemistry: Elementary band representations. *Phys. Rev. B* **2018**, 97, 035139.

(35) Hu, X.; Wu, Z.; Li, Z.; Xu, Q.; Chen, K.; Hu, W.; Ni, Z.; Jin, K.; Weng, H.; Lu, L. High-throughput search for lossless metals. *Phys. Rev. Mater.* **2022**, 6, 065023.

(36) Kresse, G.; Furthmuller, J. Efficient iterative schemes for ab initio total-energy calculations using a plane-wave basis set. *Phys. Rev. B* **1996**, 54, 11169-11186.

(37) Perdew, J. P.; Burke, K.; Ernzerhof, M. Generalized Gradient Approximation Made Simple. *Phys. Rev. Lett.* **1996**, 77, 3865-3868.

(38) Heyd, J.; Scuseria, G. E.; Ernzerhof, M. Hybrid functionals based on a screened Coulomb potential. *J. Chem. Phys.* **2003**, 118, 8207-8214.

(39) Heyd, J.; Scuseria, G. E.; Ernzerhof, M. Erratum:"Hybrid functionals based on a screened coulomb potential" [J. Chem. Phys. 118, 8207 (2003)]. *J. Chem. Phys.* **2006**, 124, 219906.

(40) Blum, V.; Gehrke, R.; Hanke, F.; Havu, P.; Havu, V.; Ren, X.; Reuter, K.; Scheffler, M. Ab initio molecular simulations with numeric atom-centered orbitals. *Comput. Phys. Commun.* **2009**, 180, 2175-2196.

(41) Ren, X.; Rinke, P.; Blum, V.; Wieferink, J.; Tkatchenko, A.; Sanfilippo, A.; Reuter, K.; Scheffler, M. Resolution-of-identity approach to Hartree – Fock, hybrid density functionals, RPA, MP2 and GW with numeric atom-centered orbital basis functions. *New J. Phys.* **2012**, 14, 053020.

(42) Levchenko, S. V.; Ren, X.; Wieferink, J.; Johanni, R.; Rinke, P.; Blum, V.; Scheffler, M. Hybrid functionals for large periodic systems in an all-electron, numeric atom-centered basis framework. *Comput. Phys. Commun.* **2015**, 192, 60-69.

(43) Tersoff, J.; Hamann, D. R. Theory and application for the scanning tunneling microscope. *Phys. Rev. Lett.* **1983**, 50, 1998-2001.

(44) Gao, J.; Qian, Y.; Jia, H.; Guo, Z.; Fang, Z.; Liu, M.; Weng, H.; Wang, Z. Unconventional materials: the mismatch between electronic charge centers and atomic positions. *Sci Bull* **2022**, 67, 598-608.

(45) Xu, Y.; Elcoro, L.; Song, Z.; Vergniory, M. G.; Felser, C.; Parkin, S. S.; Regnault, N.; Manes, J. L.; Bernevig, B. A. Filling-Enforced Obstructed Atomic Insulators. 17 Jun 2021. arXiv. https://arxiv.org/abs/2106.10276.

(46) He, Y.; Jiang, Y.; Zhang, T.; Huang, H.; Fang, C.; Jin, Z. SymTopo: An automatic tool for calculating topological properties of nonmagnetic crystalline materials. *Chin. Phys. B* **2019**, 28, 087102.

(47) Bradlyn, B.; Elcoro, L.; Cano, J.; Vergniory, M. G.; Wang, Z.; Felser, C.; Aroyo, M. I.; Bernevig, B. A. Topological quantum chemistry. *Nature* **2017**, 547, 298-305.

(48) Lax, M. Symmetry principles in solid state and molecular physics. Courier Corporation, **2001.**

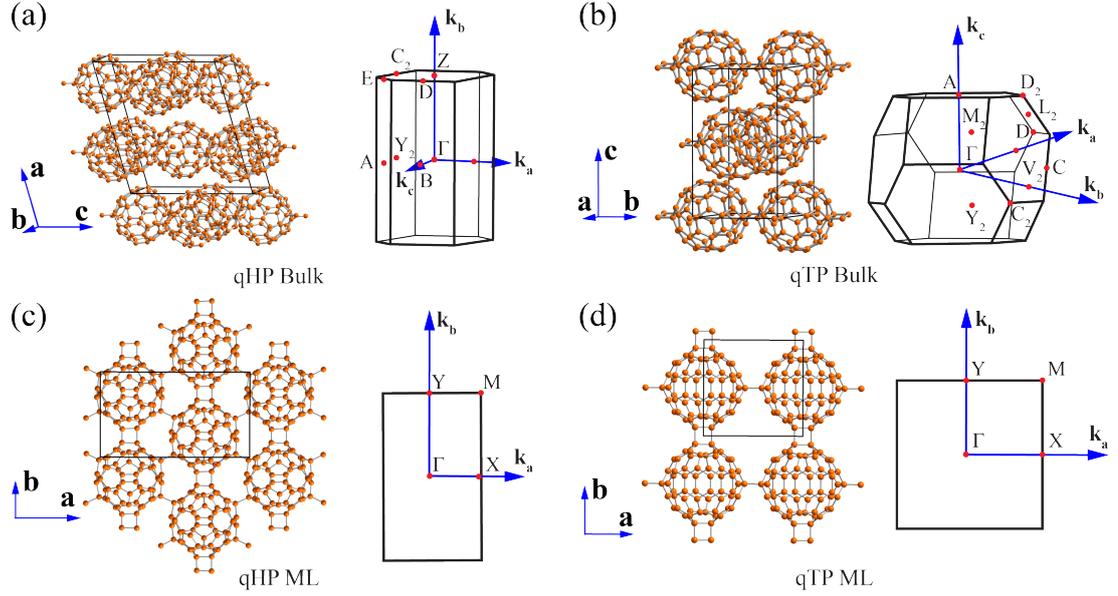

**Figure 1**. Crystal structures and the corresponding first Brillion zones (BZs) for: (a) qHP Bulk, (b) qTP Bulk, (c) qHP ML, (d) qTP ML.

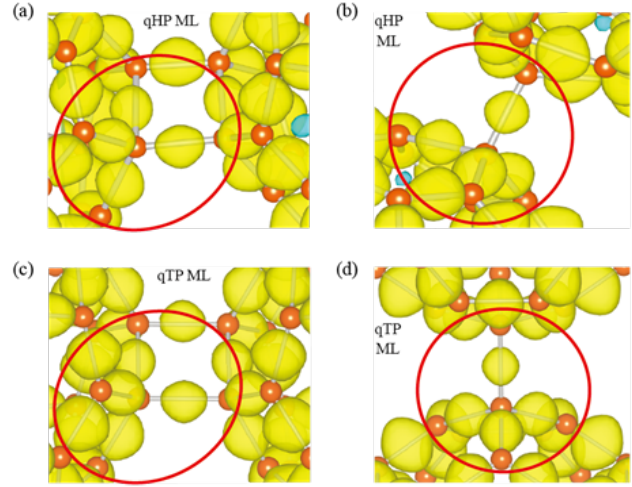

**Figure 2**. The charge difference density around C-C bonds connecting adjacent $C_{60}$ clusters in qHP ML and qTP ML. (a) Bonds in square ring and (b) the single bond in qHP ML. (c) Bonds in square ring and (d) the single bond in qTP ML. The isosurface value is set as 0.022 electrons/Å$^3$ for qHP ML and 0.024 electrons/Å$^3$ for qTP ML. The yellow and cyan represent electron and hole, respectively.

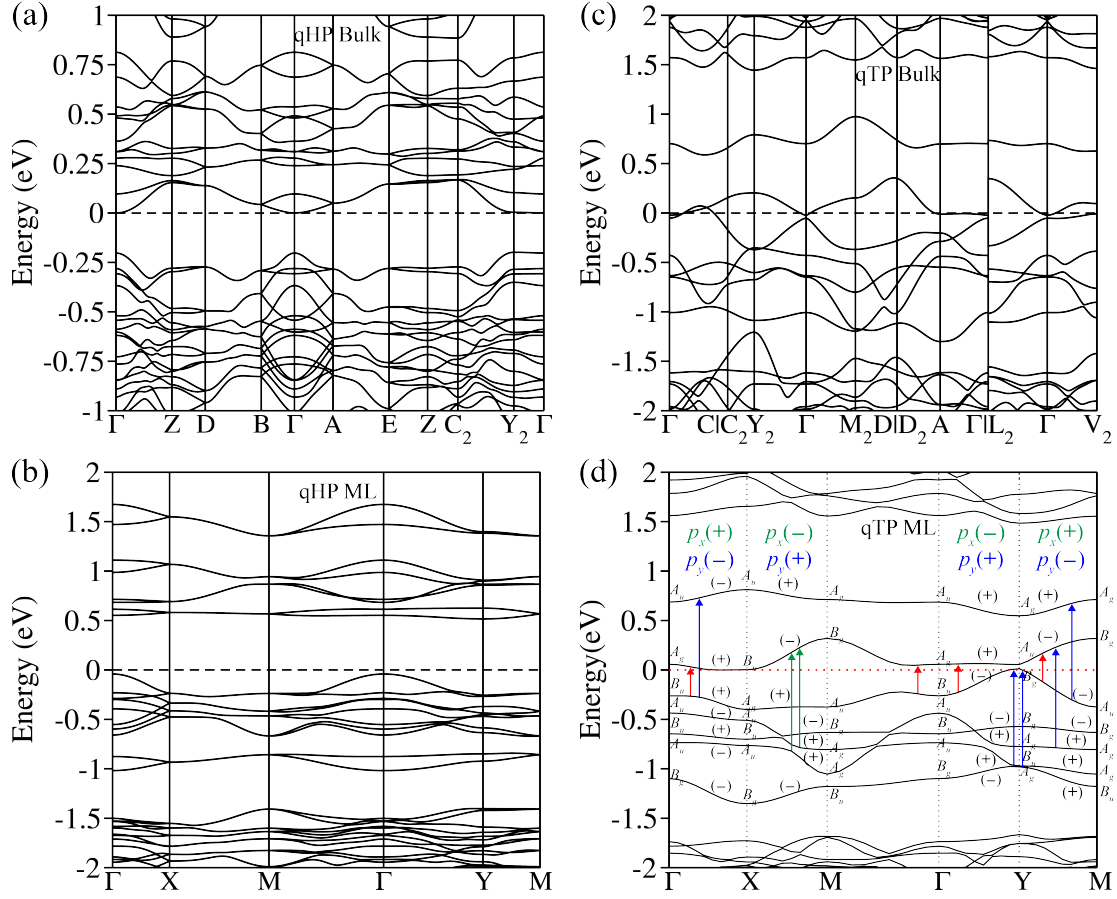

**Figure 3**. Band structures calculated with PBE functional for (a) qHP Bulk, (b) qHP ML, (c) qTP Bulk, (d) qTP ML. In (d), bands at the high-symmetry points are marked with their irreducible representations (IRREPs) in $C_{2h}$ group. Along the high symmetry lines Γ-Y, X-M (with $C_{2y}$ rotation symmetry) and Γ-X, Y-M (with $M_y$ mirror symmetry), the bands and momentum operator ($\hat{p}_x$, $\hat{p}_y$) that are even (odd) under $C_{2y}$ or $M_y$ are marked with + (−). The red arrows represent the interband transitions at the photon energy of 0.29 eV in Figure 5(b). The blue and green arrows represent those excited by x-polarized and y-polarized incident light, respectively, with the photon energy ranging from 0.95 to 1.05 eV in Figure 5(b).

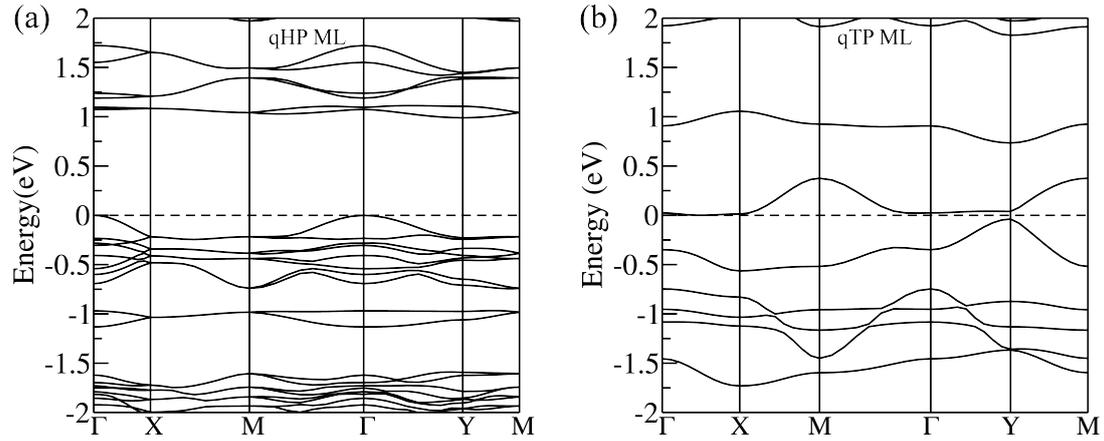

**Figure 4**. Band structures using HSE06 hybrid functional for (a) qHP ML, (b) qTP ML.

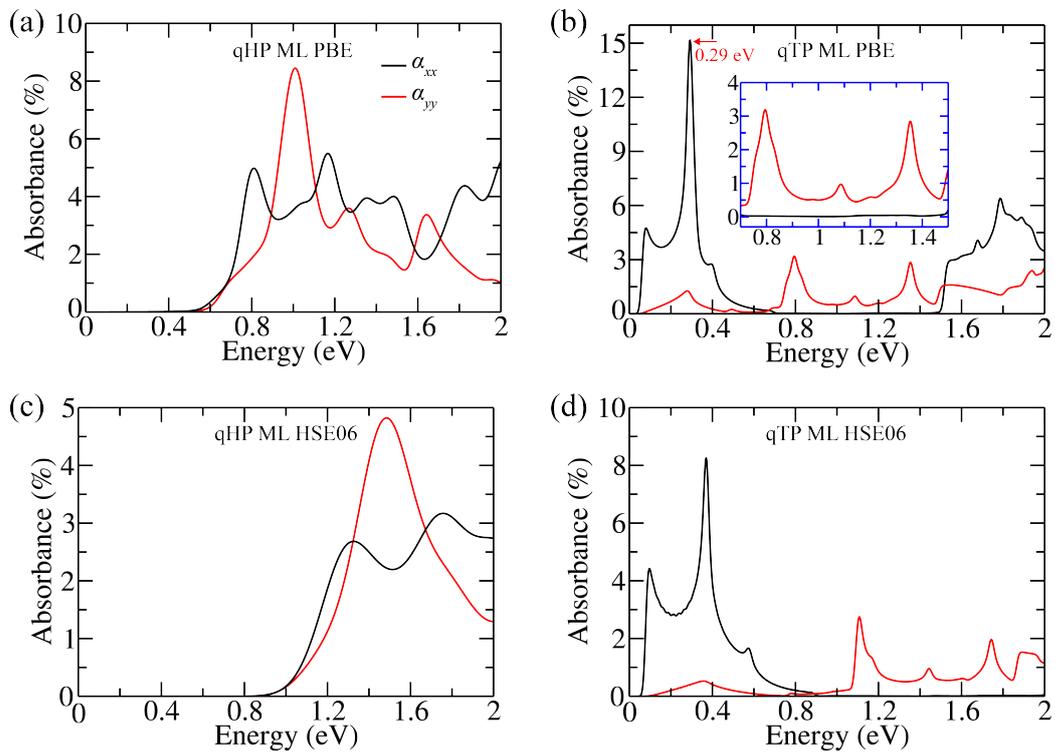

**Figure 5**. Calculated optical absorption spectra using PBE functional for: (a) qHP ML and (b) qTP ML. Those calculated with HSE06 functional are in (c) qHP ML and (d) qTP ML. The black lines represent the incident light with polarization along *x* direction and the red lines stand for those along *y* direction.

**Table 1.** Character table and basis functions of point group (PG) $C_{2h}$, $C_2$, and $C_s$. The basis function $x/y/z$ ($R_{x/y/z}$) denotes the polar (axial) vector along $x/y/z$ direction.

| PG | IRREP | E | $C_{2y}$ | P | $M_y$ | Basis function |
|---|---|---|---|---|---|---|
| $C_{2h}$ | $A_g$ | 1 | 1 | 1 | 1 | $1, R_y$ |
| | $A_u$ | 1 | 1 | −1 | −1 | $y$ |
| | $B_g$ | 1 | −1 | 1 | −1 | $R_x, R_z$ |
| | $B_u$ | 1 | −1 | −1 | 1 | $x, z$ |
| $C_2$ | $A$ | 1 | 1 | | | $1, y, R_y$ |
| | $B$ | 1 | −1 | | | $x, z, R_x, R_z$ |
| $C_s$ | $A'$ | 1 | | 1 | | $1, x, z, R_y$ |
| | $A''$ | 1 | | | −1 | $y, R_x, R_z$ |

**Table 2.** The irreducible representations (IRREPs) at high-symmetry points (HSPs) and corresponding elementary band representations (EBRs) of the qTP ML. The notation of EBR, i.e., D@W, means an orbital of IRREP D is placed at Wyckoff position W, and generates the corresponding IRREPs in the BZ. $N$ denotes the number of occupied bands and in this work the $N$-th band at Γ corresponds to the highest occupied molecular orbital (HOMO) and the $(N+1)$-th corresponds to the lowest unoccupied molecular orbital (LUMO).

| Band | Γ(0,0) | X(0.5,0) | Y(0,0.5) | M(0.5,0.5) | EBR |
|---|---|---|---|---|---|
| $N+2$ | $A_u$ | $A_u$ | $A_g$ | $A_g$ | $A_u$@(0,0.5) |
| $N+1$ | $A_g$ | $B_u$ | $A_u$ | $B_g$ | $A_g$@(0.5,0.5) |
| $N$ | $B_u$ | $A_g$ | $B_g$ | $A_u$ | $B_u$@(0.5,0.5) |
| $N-4 \sim N-1$ | $2A_u, B_g, B_u$ | $2A_u, B_g, B_u$ | $2A_g, B_g, B_u$ | $2A_g, B_g, B_u$ | $2A_u@(0,0.5), B_g, B_u@(0,0)$ or $2A_u, B_g, B_u@(0,0.5)$ |

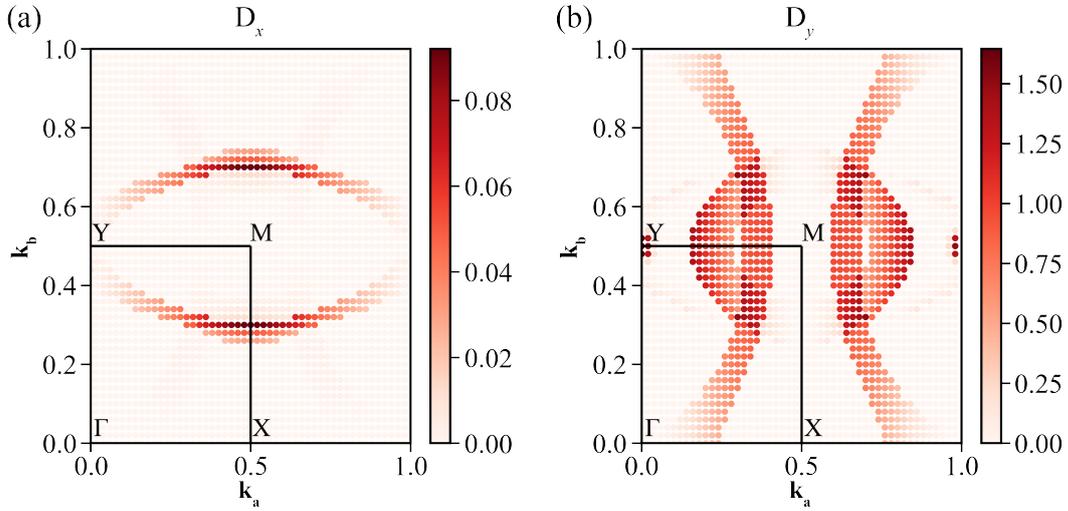

**Figure 6.** The distribution of dipole transition momentum strength (a) $D_x$ and (b) $D_y$ with the incident photon energies between 0.95 eV and 1.05 eV in the BZ of qTP ML.

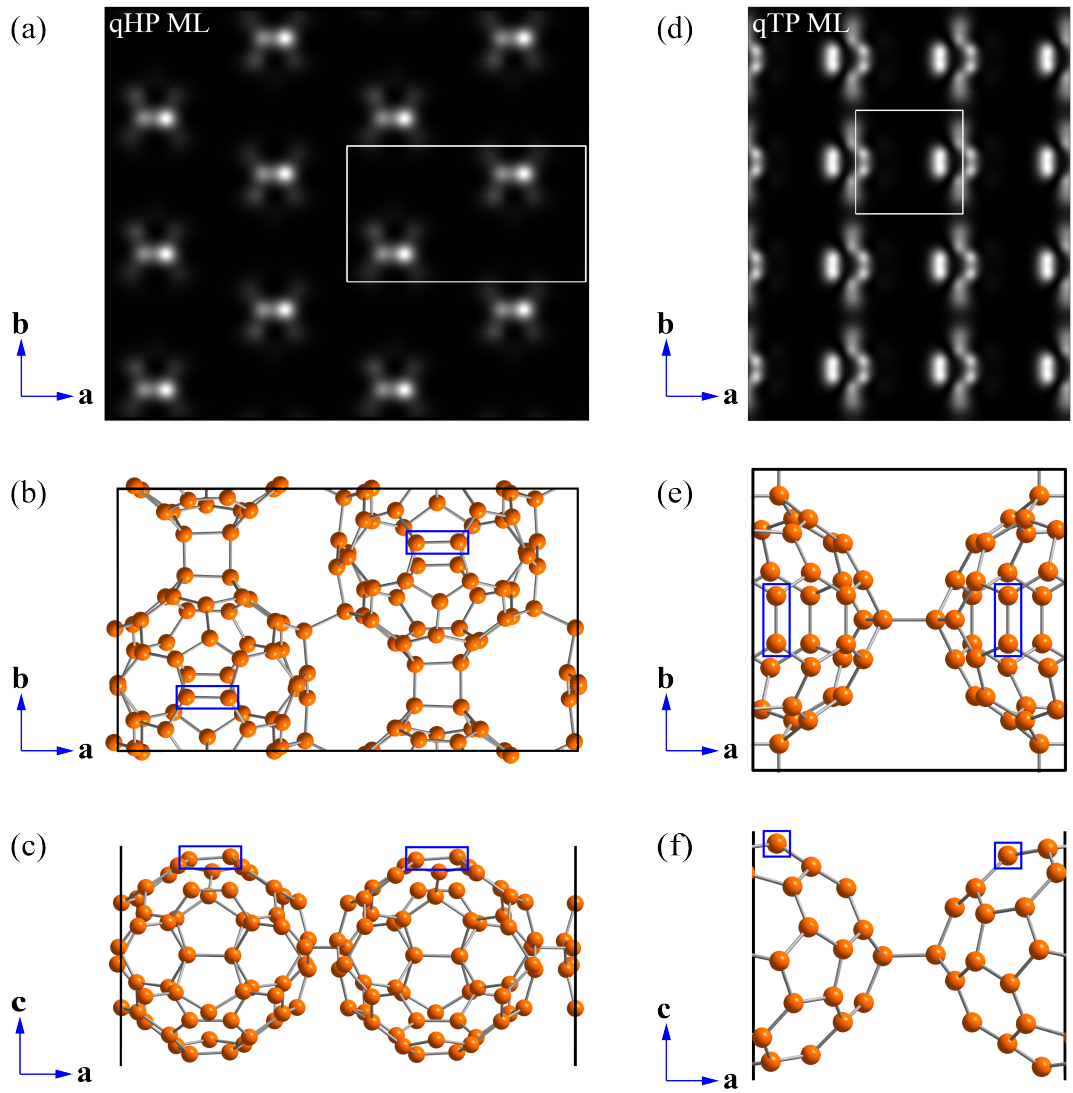

**Figure 7**. (a) Calculated STM image of qHP ML, and integration is done in the energy range from -1.0 to 0.0 eV related to Fermi energy $E_f$. The crystal structure of primitive cell for qHP ML: (b) top view, (c) side view. (d) Calculated STM image of qTP ML, and the integration is performed in the energy range from -0.1 eV to 0.1 eV related to $E_f$. The crystal structure of primitive cell for qTP ML: (e) top view, (f) side view. The distance between tip and ML is set to be 1 Å. The white rectangles in STM images represent the primitive cells of qHP ML and qTP ML. The blue frames label the corresponding C-C bonds highlighted in the STM images.